\documentclass[aps,reprint,superscriptaddress,english]{revtex4-1}

\usepackage{hyperref}
\usepackage{amsmath}
\usepackage{amsfonts}
\usepackage{amssymb}
\usepackage{graphicx}
\usepackage{array}
\usepackage{adjustbox}
\usepackage{xcolor}
\usepackage{todonotes}
\usepackage{physics}

\newcommand{\YSO}{Y$_2$SiO$_5$}
\newcommand{\ErYSO}{Er$^{3+}$:Y$_2$SiO$_5$}
\newcommand{\pErYSO}{$^{167}$Er$^{3+}$:Y$_2$SiO$_5$}
\newcommand{\gnd}{$^4$I$_{15/2}$}
\newcommand{\ex}{$^4$I$_{13/2}$}

\newcommand{\pEr}{$^{167}$Er$^{3+}$}

\begin{document}

\title{Long spin coherence times in the ground state and an optically excited state of \pErYSO\ at zero magnetic field}

\author{Jelena V. Rakonjac}
\altaffiliation[Present address: ]{ICFO-Institut de Ciencies Fotoniques, The Barcelona Institute of Technology, Mediterranean Technology Park, 08860 Castelldefels (Barcelona), Spain}
\email{j.v.rakonjac@gmail.com}
\affiliation{The Dodd-Walls Centre for Photonic and Quantum Technologies, New Zealand}
\affiliation{Department of Physics, University of Otago, Dunedin 9016, New Zealand}
\author{Yu-Hui Chen}
\affiliation{The Dodd-Walls Centre for Photonic and Quantum Technologies, New Zealand}
\affiliation{Department of Physics, University of Otago, Dunedin 9016, New Zealand}
\affiliation{School of Physics, Beijing Institute of Technology, 5 South Zhongguancun Street, Beijing 100081, China}
\author{Sebastian P. Horvath}
\affiliation{The Dodd-Walls Centre for Photonic and Quantum Technologies, New Zealand}
\affiliation{Department of Physics, University of Otago, Dunedin 9016, New Zealand}
\affiliation{Department of Physics, Lund University, P.O. Box 118, SE-22100 Lund, Sweden.}
\author{Jevon J. Longdell}
\email{jevon.longdell@otago.ac.nz}
\affiliation{The Dodd-Walls Centre for Photonic and Quantum Technologies, New Zealand}
\affiliation{Department of Physics, University of Otago, Dunedin 9016, New Zealand}

\date{\today}

\begin{abstract}
Spins in solids are an ideal candidate to act as a memory and interface with superconducting qubits due to their long coherence times. We spectroscopically investigate erbium-167-doped yttrium orthosilicate as a possible microwave-addressed memory employing its microwave frequency transitions that occur without applying an external magnetic field. We obtain coherence times of 380\,$\mu$s in a ground state spin transition and 1.48\,ms in an excited state spin transition. This is 28 times longer compared to previous zero field measurements, as well as 200 times longer than a previous microwave memory demonstration in the same material. These long coherence times show that erbium-167-doped yttrium orthosilicate has potential as a microwave-addressed quantum memory.
\end{abstract}

\pacs{}

\maketitle


\section{Introduction}

Among the wide range of physical systems presently under investigation for quantum information applications, superconducting qubits are one of the most promising. They have achieved both fast gate operations as well as very promising scalability \cite{xiang2013a,krantz_quantum_2019}. However, to date, superconducting systems face two key shortcomings: They have a limited coherence time, and, since they are addressed using microwave frequency photons, they need to operate at millikelvin temperatures, which inhibits long distance communication. The field of hybrid quantum computing aims to resolve these limitations by interfacing superconducting qubit devices with physical systems that exhibit long coherence times and/or can be addressed with optical photons. A number of physical implementations have been proposed for performing these ancillary functions \cite{lambert2020}, including optomechanical resonators \cite{andr14,bagc14,boch13}, electro-optic devices \cite{rued16,fan_superconducting_2018}, Rydberg atoms \cite{kiff16,han_coherent_2018,vogt_efficient_2019,covey_microwave--optical_2019}, nitrogen-vacancy centers \cite{kubo2011,kubo2012,grezes2014,grezes2015}, magnonic excitations \cite{hisa16,zhan16,haig16} and spins in fullerene cages and semiconductors~\cite{wu2010}.

Rare-earth ion doped crystals have been one of the physical systems at the forefront of hybrid quantum computing with superconducting circuits, with implementations of microwave memories \cite{probst2015a,wolfowicz2015}, atomic spin-ensemble upconversion using rare earths in solids \cite{will14b,obrien2014,fernandez-gonzalvo_cavity-enhanced_2019,bartholomew2019} along with a proposal for magnon based upconversion \cite{everts_microwave_2019}. The rare-earth erbium is particularly attractive as it has an optical transition in the 1.5 $\mu$m telecommunications band. Naturally occurring erbium has six even isotopes with no nuclear spin and one odd isotope, erbium-167 (\pEr), with a nuclear spin of 7/2, leading to a total of 16 ground-state levels. This endows erbium with an exceptional microwave bandwidth, with transition frequencies up to $\sim 6$ GHz with zero applied magnetic field. 

The interest in rare-earth ion doped crystals for other quantum information applications \cite{thiel2011} is due to their long optical and spin coherence times \cite{bottger2009,zhong2015}, which would also be needed for a microwave-addressed memory or echo-based frequency upconversion for hybrid systems \cite{obrien2014,welinski2019}. The longest spin coherence times so far have been obtained by applying specific magnetic fields so that the frequency of a selected transition is insensitive to spin-flip induced magnetic field perturbations \cite{zhong2015}. Such transitions are referred to as zero first-order Zeeman (ZEFOZ) transitions in solid-state media, or `clock transitions' in atomic physics. This poses a problem for hybrid quantum systems, since the performance of superconducting qubits is impaired by the application of large external fields \cite{bothner2011}. An exception to this is the recent work by Ortu \emph{et al}. \cite{ortu2018}, which investigated ZEFOZ points at zero field in ytterbium-171 doped yttrium orthosilicate (${}^{171}$Yb$^{3+}$:\YSO) and achieved coherence times exceeding 1 ms for microwave transitions. This was possible because ytterbium is a Kramers ion (unlike Eu$^{3+}$:Y$_2$SiO$_5$ used in Ref.~\cite{zhong2015}), with an odd number of valence electrons. In a low symmetry host (like Y$_2$SiO$_5$) the $J$ multiplets for a Kramers ion split into doubly degenerate ``Kramers doublets''. Each of these doublets can be treated as having an effective spin of 1/2. The hyperfine interaction between these electronic doublets and the $^{171}$Yb nuclear spin resulted in avoided crossings and thus ZEFOZ transitions at zero field. Erbium is also a Kramers ion so for \pErYSO\ ZEFOZ transitions are also present at zero magnetic field.

Welinski \emph{et al}. \cite{welinski2019} measured the spin coherence times for the even isotopes of \ErYSO\ arriving at coherence times of the order of 1.6\,$\mu$s. The relatively short coherence times are the result of the large magnetic field sensitivity of an electron spin. Ran\v{c}i\'{c} \emph{et al}. \cite{rancic2018} achieved a remarkable hyperfine coherence time for \pErYSO\ of 1.3\,s using a 7 T magnetic field. This large coherence time was the result of two factors. First, the temperature was low enough and the field high enough to freeze out the electron spins, significantly reducing the magnetic noise. Second, at such large magnetic fields the electron spin states and the nuclear spin states are unmixed and the transitions probed were essentially nuclear spin transitions with transition dipole moments of the order of the nuclear magneton (14 MHz/T). The resulting weak oscillator strength means that these transitions are poorly suited for coupling directly to microwave photons and superconducting qubits. Operating, as we do here, close to zero magnetic field blurs the distinction between `electron spin' and `nuclear spin' transitions. Thus, transitions in this regime can potentially have long coherence times, as well as comparably large transition dipole moments.

The benefit of freezing out electron spins on spin coherence times has also been shown in recent ENDOR (electron-nuclear double resonance) measurements on $^{143}$Nd$^{3+}$:Y$_2$SiO$_5$ where 40\,ms (2\,ms) coherence times were achieved for nuclear (electron) spins at 100\,mK and $\mathcal{O}(100\,\text{mT})$ fields~\cite{li2020}.

Spin coherence times in \pErYSO\ have been investigated at zero field by Hashimoto \emph{et al}. \cite{hashimoto2016a}. The authors used coherent Raman beats and arrived at a lower bound for the coherence times of what they called a ``sublevel coherence'' of 50\,$\mu$s.

For this work we perform Raman heterodyne spectroscopy to identify hyperfine transitions in \pErYSO\ at zero field that have the potential to have long coherence times. We use the standard two pulse spin echo sequence to measure significantly longer coherence times, and extend them further using dynamic decoupling.

\section{Structure of $^{167}$E\lowercase{r}$^{3+}$:Y$_2$S\lowercase{i}O$_5$}

\begin{figure}
\includegraphics[width=\linewidth]{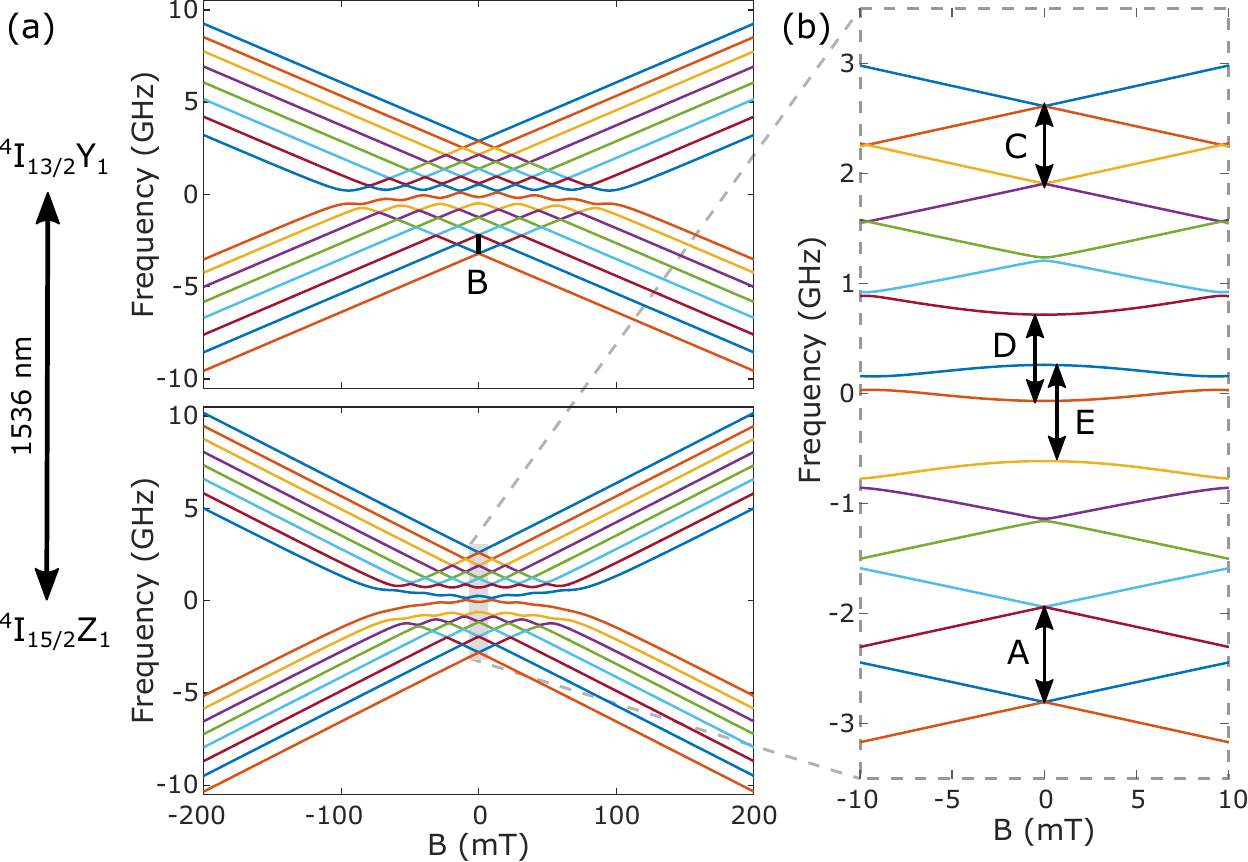}
\caption{\label{fig:lvls}(a) Energy levels of the \gnd$Z_1$ ground \cite{chen2018} and optically excited \ex$Y_1$ \cite{horvath2019} states of \pEr\ as a function of magnetic field applied along the $D_1$ axis. Adjacent levels are plotted with different colours for distinguishability. (b) More detail of the behavior in the ground state near zero magnetic field. The transitions for which we investigate coherence properties are indicated.}
\end{figure}

The host crystal, \YSO, is known as a ``low magnetic noise'' host due to the small contribution from its constituent ions to the magnetic noise (and thus dephasing) experienced by a dopant. Yttrium has one stable isotope with a small nuclear magnetic moment, and only uncommon isotopes oxygen and silicon have nuclear magnetic moments. As is convention for this material we use the coordinate system defined by the principal axes of polarization, $D_1$, $D_2$, and $b$~\cite{li1992}. \pEr\ ions can substitute for Y$^{3+}$ ions in two crystallographic sites, referred to as site 1 and site 2~\cite{macfarlane1997}. Ions in site 1 and site 2 experience different crystal fields and different optical transition frequencies. The \pEr\ ions in each of these sites can be further divided into four subsites. The four subsites are related to each other by inversion and rotation about the crystal's $C_2$ axis. At zero magnetic field all four subsites have identical energy levels. In an applied magnetic field, all subsites have identical transition frequencies for magnetic fields applied along the $b$ axis or in the $D_1$-$D_2$ plane; however, the degeneracy of pairs related by the $C_2$ rotation will be lifted and the number of lines present in a spectrum will double if a magnetic field is applied in any other direction. The degeneracy of the pairs of subsites related by inversion is not lifted by a magnetic field.

A free Er$^{3+}$ ion has a 16-fold degenerate ground state $^4$I$_{15/2}$ and 14-fold degenerate first excited state $^4$I$_{13/2}$. In \YSO\ these split into eight and seven Kramers doublets respectively. Here we use the lowest-lowest transition between these manifolds ($^4$I$_{15/2}Z_1\rightarrow ^4$I$_{13/2}Y_1$) which is very narrow because it is insensitive to phonon based population decay processes. The hyperfine coupling between these two Kramers doublets and the $I = 7/2$ nuclear spin results in 16 different energy levels for the ground and optically excited states. These energy levels are shown in Fig.~\ref{fig:lvls} as a function of magnetic field along the $D_1$ axis. The behavior of the spin states of the ground ($^4$I$_{15/2}Z_1$) and excited ($^4$I$_{13/2}Y_1$) manifold can each be described by a spin Hamiltonian of the form
\begin{equation}
H=\mu_e \mathbf{B} \cdot \mathbf{g} \cdot \mathbf{S} + \mathbf{I} \cdot \mathbf{A} \cdot \mathbf{S} +\mathbf{I} \cdot \mathbf{Q} \cdot \mathbf{I} -\mu_n g_n \mathbf{B} \cdot \mathbf{I} ,\label{eq:spinHam}
\end{equation} 
where $\mu_e$ is the Bohr magneton, $\mathbf{B}$ the applied magnetic field, $\mathbf{g}$ the Zeeman $g$ matrix, $\mathbf{A}$ the hyperfine matrix, $\mathbf{Q}$ the electric quadrupole matrix, $\mu_n$ the nuclear magneton, and $g_n = -0.1618$ is the nuclear $g$ factor. The spin Hamiltonian for the ground state has been determined by electron spin resonance experiments \cite{chen2018} and a spin Hamiltonian for the excited state can be derived from a recent crystal field model \cite{horvath2019}.

The hyperfine coupling is very approximately of the form $H = AI_zS_z$, and a Hamiltonian of that form leads to eight pairs of hyperfine energy levels that cross at zero magnetic field. The perturbation to this situation caused by the rest of the hyperfine coupling terms turns these crossings all into avoided crossings. As shown in Fig.~\ref{fig:lvls}(b), for the highest and lowest energy level pairs of the hyperfine manifold, the size of the anticrossing is small and for most purposes it can be considered that the energy levels cross. For energy levels in the middle of the manifold the anticrossing is large leading to single energy levels whose energies are insensitive to a magnetic field to first order. This results in two different types of transitions with low magnetic field sensitivity and thus the potential for long coherence times. The first is those that are between levels with similar magnetic sensitivity, such as A, B, and C in Fig.~\ref{fig:lvls}. We call these doublet-doublet transitions. The second type are ZEFOZ transitions such as D in Fig.~\ref{fig:lvls}.

\section{Experimental Setup}

\begin{figure}
\includegraphics[width=\linewidth]{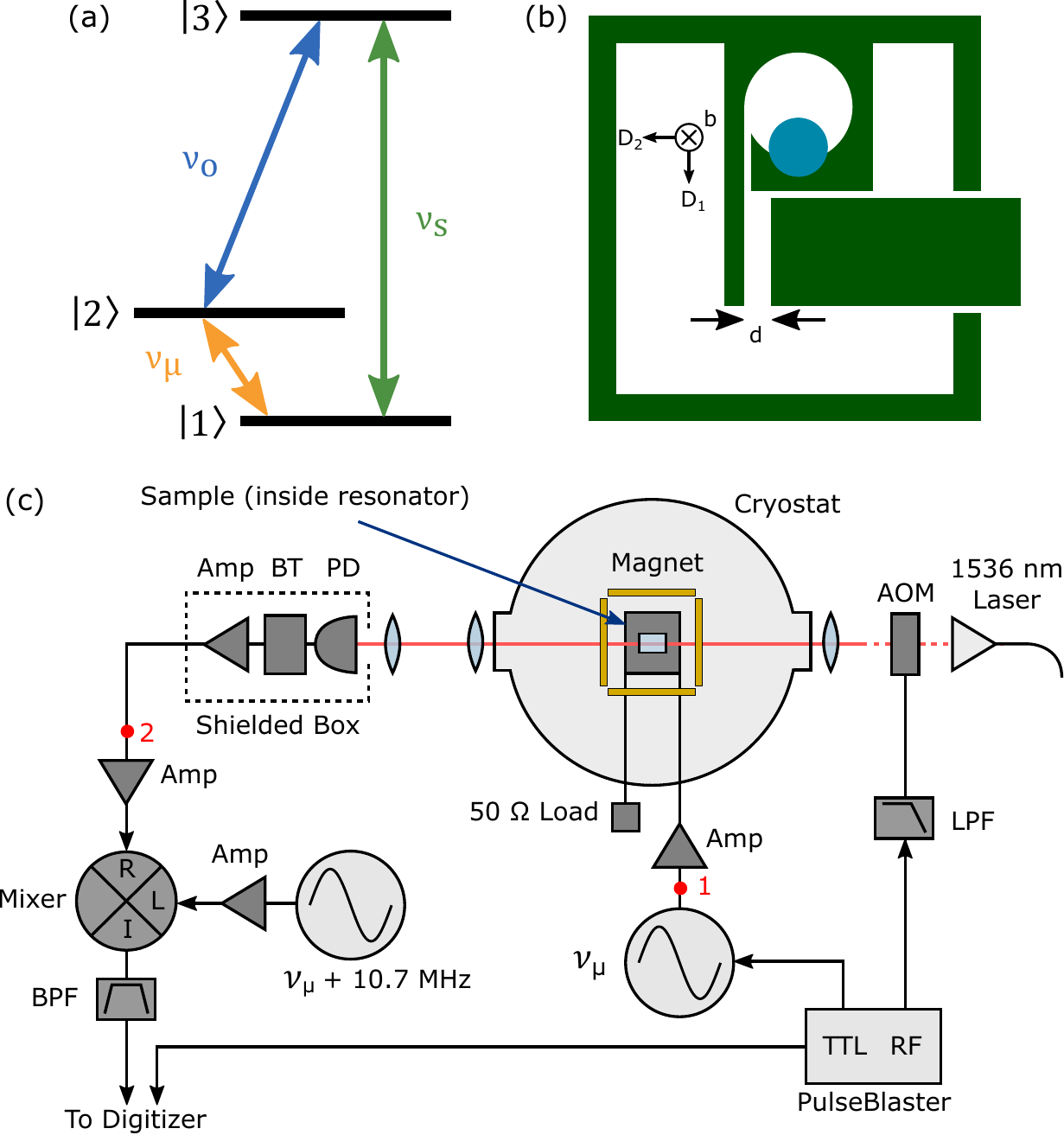}
\caption{\label{fig:setup} (a) Level scheme for Raman heterodyne measurements. (b) Cross section of resonator (green) with the sample (blue) inside. The orientation of the $D_1$, $D_2$, and $b$ axes are indicated on the side. (c) Setup diagram for pulsed Raman heterodyne measurements (see main text for acronym definitions). For CW measurements, most of the setup is bypassed and a network analyzer is placed between points 2 and 1, where it sends a probing signal at point 1 and receives it at point 2.}
\end{figure}

To measure hyperfine spectra and coherence times, we use Raman heterodyne spectroscopy~\cite{mlynek1983,wong1983}. This method allows hyperfine transitions at microwave frequencies to be detected optically. Figure~\ref{fig:setup} (a) shows the level scheme for this process, where levels $\ket{1}$ and $\ket{2}$ are two hyperfine levels in the ground state and $\ket{3}$ is a hyperfine level in the excited state. By driving the hyperfine transition between $\ket{1}$ and $\ket{2}$ with a microwave field of frequency $\nu_\mu$ and the optical transition between $\ket{2}$ and $\ket{3}$ with a laser at frequency $\nu_o$, another optical field at the sum of these two frequencies, $\nu_s$, will be generated. The two optical frequencies will beat together with a frequency of $\nu_\mu$, which can be detected optically. This method also works in the case where $\ket{1}$ and $\ket{2}$ are excited-state hyperfine transitions and $\ket{3}$ is in the ground state, so hyperfine transitions in both the ground and excited states can be detected simultaneously.

We use a cylindrical \pErYSO\ sample (Scientific Materials Inc.) with a length of 12.0\,mm and a diameter of 4.95\,mm. 50 parts per million of the Y$^{3+}$ ions are substituted for \pEr\ ions. Our laser frequency is chosen to only address ions in site 1.

The sample is placed inside an aluminium, tunable loop-gap microwave resonator [Fig.~\ref{fig:setup} (b)] based on a single-loop, single-gap design \cite{hardy1981}. The resonator was placed inside a homebuilt cryostat (cooling head: Cryomech PT405) containing a three axis superconducting magnet (HTS-110 Ltd.), which allows us to apply a magnetic field in an arbitrary direction. The $^4I_{15/2}$ to $^4I_{13/2}$ transition in site 1 is driven optically along the crystal's $b$ axis with a fiber laser (Koheras E15 DFB). The polarization used was not well characterized, but we used waveplates to adjust the unknown polarization of the laser to maximize the Raman heterodyne signal. The laser power was typically 20\,mW before the cryostat.

Like our designs for higher frequencies \cite{chen2018}, the resonant frequency is tuned by adjusting the gap $d$ with a plunger attached to a piezoactuator (ANPx101, attocube systems). The loop was made bigger than the sample to get a resonant frequency tunable in the range of about 700\,MHz to 1200\,MHz. At resonance, the radio frequency (RF) magnetic field oscillates in and out of the loop, which is along the $b$ axis of the crystal. Small loops of wire are placed concentric to the main resonator loop to couple RF into the resonator.

To perform the CW Raman heterodyne measurements, we use microwaves generated by a network analyzer (FieldFox) which are amplified to give a typical power of around 50 mW to drive hyperfine transitions in \pErYSO\ inside the resonator [Fig.~\ref{fig:setup}(c) between points 1 and 2]. The resonator is kept at a fixed frequency for these measurements as its linewidth is sufficiently broad such that transitions a few hundred megahertz away from the resonant frequency can be detected.

The optical signal with Raman heterodyne modulation is measured with a fast photodetector (PD). The detector is connected to a bias tee (BT), which allows the photodetector to be biased by a battery connected to the DC input, and the modulated signal to be retrieved from the RF output. The photodetector, bias tee, and two amplifiers are all enclosed in a shielded aluminium box which also contains agar to minimize RF pickup noise entering the detection chain. The Raman heterodyne signal passes through +42\,dB of amplification before reaching the network analyzer. The measured spectrum is typically averaged 10 times to suppress the background noise.

The setup for the pulsed experiments (for measuring coherence times) is shown in Fig.~\ref{fig:setup}(c), where instead of the network analyzer, a pulsed microwave source is used, and the resulting Raman heterodyne signal is analyzed in the time domain by employing a chain of frequency mixers to obtain a 10.7\,MHz signal suitable for digitization. The signal from the microwave source is amplified before going to the resonator, giving a power on the order of 10 W. The microwave source as well as the AOM which gates the laser is controlled by a PulseBlaster.

The resulting Raman heterodyne signal is amplified by +74\,dB before being mixed to a 10.7\,MHz signal through the use of a microwave source operating 10.7\,MHz higher than the frequency of the transition ($\nu_\mu$) being detected.

All of the signal sources are phase locked and the wait time between measurements was chosen carefully so that each echo signal always appears with the same phase. This allows the echo signals to be coherently averaged. The signals are typically averaged 100 to 200 times depending on the signal strength. The pulse sequences themselves are cycled every 10\,Hz for the two pulse spin echo measurements or 5\,Hz for dynamic decoupling measurements to prevent the large number of high-power microwave pulses from heating the resonator.

\section{Results}
\subsection{Spectroscopy}

\begin{figure}
\includegraphics[width=\linewidth]{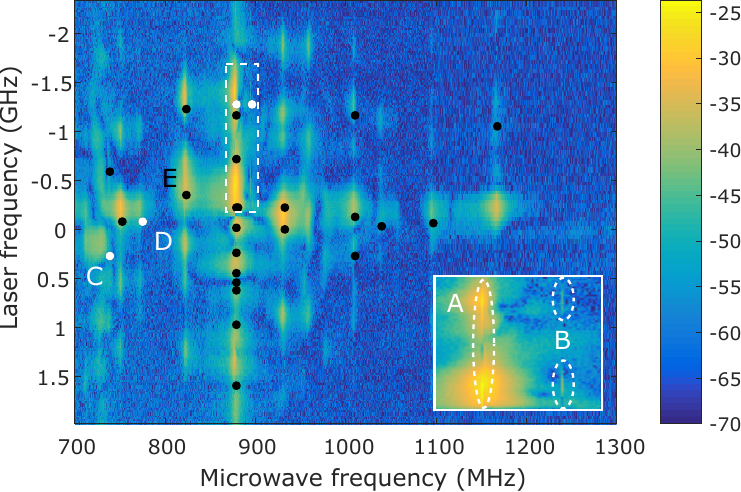}
\caption{\label{fig:rhzero} Raman heterodyne signal as a function of laser and microwave frequencies. Laser frequency is given by the detuning from 195116.5 GHz. Color scales logarithmically with signal intensity (in dB). The dots indicate transitions investigated with applied magnetic fields including transition E which is discussed further in the text. White dots indicated transitions from which spin echoes were measured (A, B, C, and D). The inset shows an enlarged view of the particularly narrow transitions A and B.}
\end{figure}

We searched for transitions by first sweeping the applied optical and microwave frequencies at zero field (the resonator frequency was kept fixed). This resulted in the 2D map of transitions shown in Fig.~\ref{fig:rhzero}, made up of three sweeps of 200 MHz in microwave frequency each. Here, regions of higher Raman heterodyne signal intensity seen in yellow and orange indicate the presence of hyperfine transitions. In general, the linewidth of many of these transitions are on the order of MHz. However, some transitions are narrower, with linewidths of 100s of kHz, two of which are noted as A and B in the inset.

To determine which transitions could yield long coherence times, we measured Raman heterodyne spectra as a function of microwave frequency and magnetic field. The transitions which we measured with an applied field are indicated with dots in Fig.~\ref{fig:rhzero}. We identified four transitions which we could measure coherence times from, noted as A, B, C, and D. A fifth transition E, from which we could not measure coherence times, is also indicated here.

The Raman heterodyne spectra for the four noted transitions are shown in Fig.~\ref{fig:compare}. In the spectra for A, four spectral lines can be seen centered around 880 MHz: two at zero field at 879.4\,MHz that appear to be degenerate and two others that appear to overlap at a slightly higher frequency, forming a kitelike shape around zero field. B and C also have two spectral lines each which appear to be degenerate at zero field, at 896.7\,MHz and 739\,MHz. However, D is a ZEFOZ transition, with a zero field frequency of 774\,MHz.

Note that both transitions A and C have been previously reported, albeit as just one transition each \cite{baldit2010}.

The double lines for A, B, and C are not due to the two magnetically inequivalent subsites and magnetic field misalignment. When the magnetic field was either applied along or perpendicular to the $C_2$ axis the double lines were observed. But magnetic fields applied in other directions caused four lines to appear. Thus we ascribe A, B and C to the doublet-doublet transitions as noted in Sec. II. For this reason, we refer to A in plural as transitions A to highlight the degeneracy and likewise for B and C. This is investigated further in Section IV. B.

\subsection{Identification of states and comparison with spin Hamiltonians}

\begin{figure}
\includegraphics[width=\linewidth]{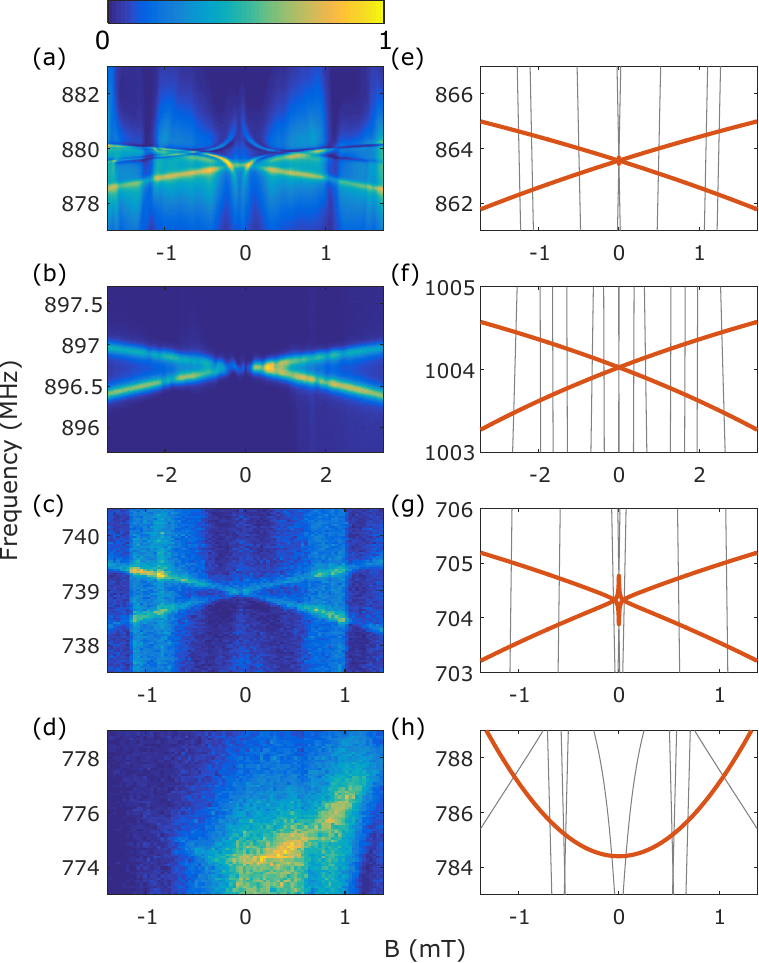}
\caption{\label{fig:compare} Comparison of Raman heterodyne spectra [(a) to (d)] and corresponding predictions from (e), (f), and (h) the ground-state spin Hamiltonian parameters from~\cite{chen2018} and (g) excited-state parameters~\cite{horvath2019}. Color scales linearly with signal intensity for (a) to (d), normalized to the minimum and maximum intensities given by the colo rbar. For (a), (c), and (d), $B || D_2$, and for (b), $B || b$, and likewise for the corresponding theory plots. In (e) to (f), bold red lines indicate the predicted transition that corresponds to the measured transition, while black lines indicate other predicted transitions in the same field and frequency range. Note the $y$-axis values differ between the Raman heterodyne data and corresponding theory plot, but are on the same scale.}
\end{figure}

Regarding the electronic state of origin of the transitions, we can determine the origin of the transitions in two ways. One way is by analyzing the predictions of the relevant spin Hamiltonians, which we will discuss first, and the other involves getting information about the electronic state lifetime using spin echoes. We note that A and C were shown to originate in the ground state in Ref. \cite{baldit2010}.

We compared the experimental spectra to the predicted spectra from the spin Hamiltonian for both the ground and excited states, and compared the behavior of the transitions in a magnetic field as well as the absolute transition frequencies. By inspection, we found the predicted line that best matches the data, and thus used the prediction to determine which hyperfine levels the transition comes from.

Figure~\ref{fig:compare} shows a comparison between the experimentally measured spectra of transitions A to D (a)-(d) and the corresponding predicted spectra (e)-(h). The predicted spectra (e)-(h) contain many spectral lines, where the lines corresponding to the transitions in the data are denoted by the bold, red lines. The $y$ axes have been chosen so that the scale remains the same as the data counterpart but shifted so that the predicted transition in bold can be seen. Many other lines, corresponding to transitions with slopes that are nearly vertical (and thus change frequency rapidly with field), are also predicted. There are some nearly-vertical features present in the experimental data [such as the lines of minimum signal intensity at $\pm$ 1 mT in Fig.~\ref{fig:compare}(a)] which could originate from these steep transitions. The kitelike pair of transitions in (a) appears in the excited-state predictions, but is not included here. There is also a kite-shaped feature in Fig.~\ref{fig:compare}(g) which was not observed in the experimental data in Fig.~\ref{fig:compare}(c).

The correspondence between the experimentally measured transitions A to D and the predictions from the ground state is very good, with the largest difference between measured and predicted zero-field frequency being 35\,MHz for C. On the scale of the entire zero-field transition frequencies (over 5\,GHz), 35\,MHz is less than 0.7\%, and this discrepancy is well within acceptable error of the ground-state parameters from Ref.~\cite{chen2018}. The discrepancy is larger for B at just over 100\,MHz. Transitions B originate from the excited state and the predictions come from the crystal field model of Ref.~\cite{horvath2019} rather than a spin Hamiltonian. 

From the comparisons shown in Fig.~\ref{fig:compare}, we can identify which energy levels the transitions come from. These levels are indicated in Fig.~\ref{fig:lvls}. A, B, and C are doublet-doublet transitions as described in Sec. II. Transition D is a ZEFOZ transition. The spin Hamiltonians do not necessarily give the correct ordering of the energy levels, so the levels could be reversed such that A comes from the higher energy doublets, while C comes from the lower energy doublets. Likewise, D is indicated as the transition between levels 8 and 10, but it could be from levels 7 and 9 instead. B is marked as the transitions between the two lowest energy doublets of the excited state, but it also resembles the transitions from the highest energy doublets (predicted around 730 MHz), so its assignment is also ambiguous.

\begin{figure}
\includegraphics[width=\linewidth]{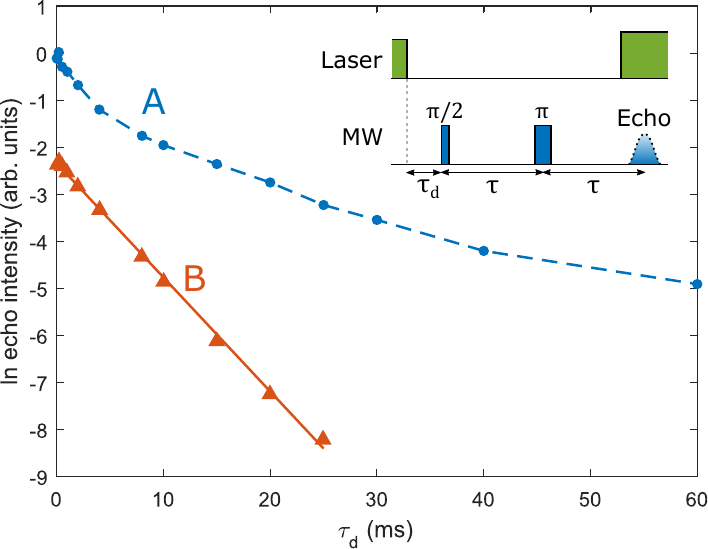}
\caption{\label{fig:dt} Dependence of echo intensity on dark time $\tau_d$ for transitions A and B at 3.2 K. A linear fitting to the echo height is plotted for transitions B, but only a guiding line between the points is shown for transition A. Inset shows the laser and microwave (MW) pulse sequence used.}
\end{figure}

We also confirmed the state of origin of some transitions using spin echoes. As per the inset in Fig.~\ref{fig:dt}, we prepare a population difference between two hyperfine levels by sending a long optical pulse at the frequency of the ground to excited-state transition. Then, some time after we turn off the laser, which we call the dark time $\tau_d$, we send a microwave $\pi$/2 and $\pi$ pulse to the resonator. We turn the laser back on at the time when we expect to observe an echo so that the echo can be detected using Raman heterodyne. By measuring the echo intensity as a function of $\tau_d$ (and keeping $\tau$ fixed), we can measure the population decay.

Figure~\ref{fig:dt} shows a plot of the (ln) echo intensities from a two pulse spin echo sequence as a function of $\tau_d$ for transitions A and B. Transitions B have a decay of 8.2\,ms, obtained from a linear fit. Transitions A do not follow a linear decay, but a rough linear fit gives a lifetime of 23\,ms. The lifetime of transitions B are consistent with the previously reported 11\,ms lifetime of the excited state \cite{bottger2006a}. Comparatively, the persistence of echoes even for $\tau_d = 60$\,ms for transitions A suggests they originate from the ground state.

Due to weak echo signals, we were not able to determine the state of origin of C and D using this method, but we are confident on the state assignment using the spin Hamiltonians.

\subsection{Coherence Measurements}

\begin{figure}
\includegraphics[width=\linewidth]{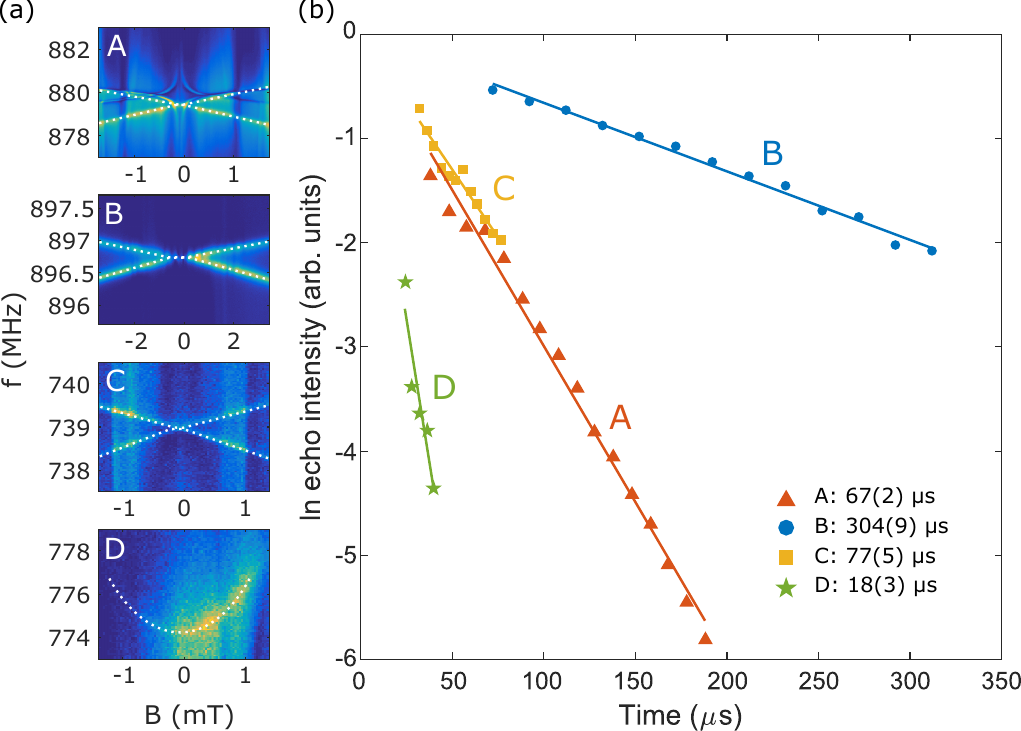}
\caption{\label{fig:T2s} Comparison of coherence times measured using two pulse spin echoes. (a) Raman heterodyne spectra as a function of magnetic field from the four transitions, with $ B \| D_2$, except for B where $B \| b$. Color scales linearly with signal intensity. White dotted guiding lines indicate the location of the particular transition and do not come from theory or fitting. (b) Intensities of the echoes measured, with corresponding linear fits, normalized using said linear fits. The resonator temperature was 3.2\,K for A and B and 3.14\,K for C and D.}
\end{figure}

We performed two-pulse spin echo measurements on the transitions A, B, C, and D. We used the same spin echo sequence as in Fig.~\ref{fig:dt}, and we used $\pi$-pulse lengths of 2\,$\mu$s for A and D, 6\,$\mu$s for B, and 4\,$\mu$s for C. 

Figure \ref{fig:T2s} shows the Raman heterodyne spectra of the transitions measured, and the echo intensity decay as a function of the time $2\tau$. To obtain the echo intensity for each measurement, we take the Fourier transform of the echo only, and fit the decay to $\exp(-2T/T_2)$, where $T = 2\tau$.

Here, B has the longest coherence time at 304\,$\mu$s, while D has the shortest. The intensity of the echoes originating from C and D were much weaker than those of A and B, hence echo measurements were made for much smaller ranges of $\tau$. There was another ZEFOZ transition, both predicted and observed (at 823.8 MHz) in the CW Raman heterodyne spectra that was flatter than D (which indicates that the coherence time should be longer \cite{fraval2004}) which we call E, but we could not detect any echo signal from it. Based on curvatures predicted from the spin Hamiltonian from Ref.~\cite{chen2018}, D and E should have yielded the longest coherence times, but instead A, B and C did; this deserves further investigation.

\begin{figure*}
\includegraphics[width=\linewidth]{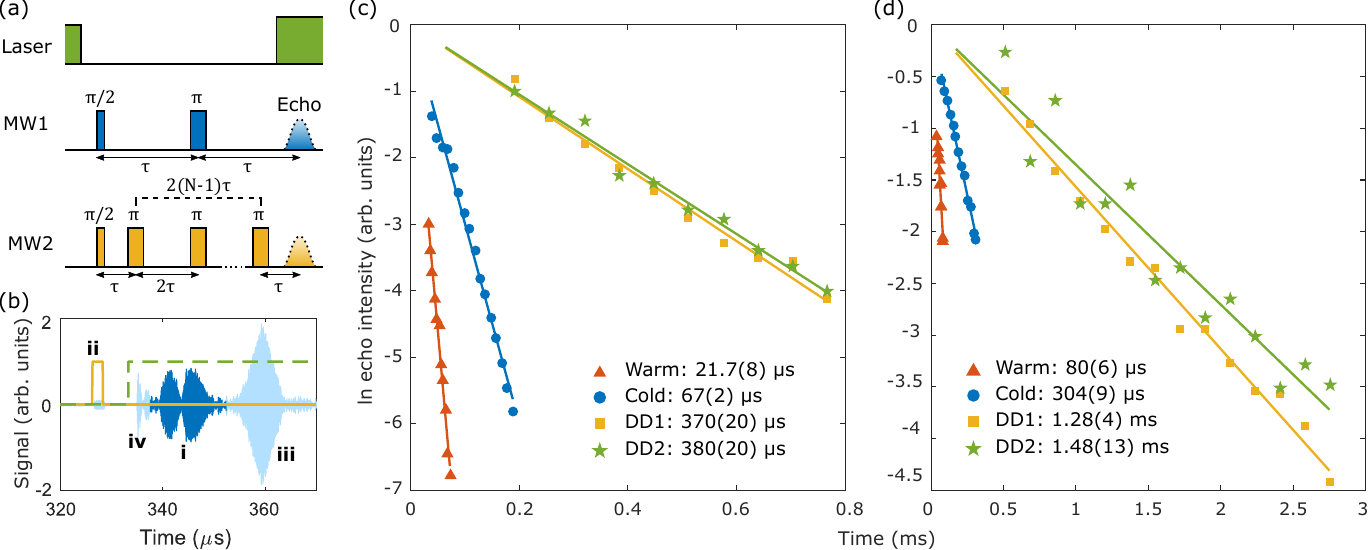}
\caption{\label{fig:echoes}(a) Pulse sequences used, where the laser sequence is the same for both MW pulse sequences: on until just before the initial $\pi/2$-pulse, then gated on again where the echo is expected to be. MW1 corresponds to a two-pulse spin echo sequence and MW2 to a dynamic decoupling pulse sequence. (b) Example of an echo from a dynamic decoupling measurement. The yellow trace indicates the timing of the final $\pi$ pulse (ii), the green dashed trace indicates the timing of the laser pulse, and the blue trace indicates the detected Raman heterodyne signal, with the main echo shown darker (i); (iii) indicates an extra echo resulting solely from the $\pi$ pulses and (iv) is leftover free induction decay picked up by the laser gating on. Echo intensity as a function of the total time $T$ since the $\pi$/2 pulse for (c) transitions A and (d) transitions B; for the two-pulse measurements, $T = 2\tau$, and for the dynamic decoupling measurements, $T = 2N\tau$ where $N$ is the number of $\pi$ pulses. `Warm' and `Cold' correspond to two-pulse spin echo measurements (MW1) made at 4.5\,K and 3.2\,K, respectively; DD1 and DD2 correspond to the two different frequency components of the echo from dynamic decoupling (MW2, see main text).}
\end{figure*}

We further investigated transitions A and B as they had longer coherence times and stronger echo signals. We could measure coherence times of A and B at a higher temperature of 4.5\,K, and we could also extend the coherence times using a dynamic decoupling pulse sequence \cite{carr1954a}. Figure~\ref{fig:echoes}(a) shows the sequence of microwave pulses used for the dynamic decoupling measurements. We kept $\tau$ fixed at 17 $\mu$s for transitions A and 46 $\mu$s for transitions B. 

In the case of the dynamic decoupling measurements, the echo envelope has a modulation which persists in the Fourier domain as two different frequency components; see Fig.~\ref{fig:echoes}(b) for an example. We fit the decay of the echo intensity to $\exp(-2T/T_2)$, where for the dynamic decoupling sequence, $T = 2N\tau$. An individual fit is made to each Fourier component of the dynamic decoupling echo. The resulting decays and fits are shown in Fig.~\ref{fig:echoes}(c) for transitions A and Fig.~\ref{fig:echoes}(d) for transitions B.

For both transitions, a decrease in temperature from 4.5\,K to 3.2\,K increased the $T_2$ by at least a factor of 3. This improvement can be expected due to the suppression of spectral diffusion processes at lower temperatures \cite{bottger2006c}. We were not able to reach lower temperatures with our cryostat, so we could not investigate this relationship further.

At 3.2\,K the dynamic decoupling pulses were applied which yielded longer coherence times still. For both transitions the echo envelope develops a modulation after a few $\pi$ pulses. Analysis of the two frequency components of the echo separately resulted in coherence times of $370 \pm 20$\,$\mu$s and $380 \pm 20$\,$\mu$s for transitions A, and $1.28 \pm 0.04$\,ms and $1.48 \pm 0.13$\,ms for transitions B. Errors are given by the $1\sigma$ standard error from the linear fitting. The values of $T_2$ for the two components as well as the combined echo are the same value within 2$\sigma$.

For all the types of measurements made, transitions B had longer coherence times than transitions A. If we consider the gradient of the transitions in a magnetic field, those of transitions A have a gradient of 350\,kHz/mT, whereas those of transitions B have a gradient of 85 kHz/mT (for $B \| b$). Therefore, the long coherence times of transitions B could be due to the transitions having a smaller first-order dependence on magnetic field. The fact that transitions B originate from the excited state can also be another reason for its longer $T_2$ as there will not be any other excited state ions to flip-flop with, aside from the ions we intentionally excite \cite{welinski2019}.

The coherence times we measure with transitions A are longer than those measured using coherent Raman beats for the same transitions~\cite{hashimoto2016a}. The authors of Ref.~\cite{hashimoto2016a} measured a coherence time of 12\,$\mu$s at the same Er concentration, and 50\,$\mu$s at a lower concentration. We measure 67\,$\mu$s using only a two pulse sequence. We also measure longer coherence times from transitions A and B compared with the microwave memory demonstrated with \ErYSO\ previously \cite{probst2015a}, where a coherence time of 5.6\,$\mu$s was measured but in the even isotopes of Er and with an applied magnetic field.

The modulation in the echo envelope suggests that the transitions A and B are not degenerate. From the spin Hamiltonian predictions, we know that A and B originate from either the lowest or highest energy hyperfine doublets and its adjacent doublet for each electronic state. The ground state and excited state spin Hamiltonians predict the separation between the doublets at either extreme of energy to be 12 kHz for one doublet and less than one kHz for the rest, so in practice the lowest and highest energy hyperfine levels should cross. The modulation could occur from stray magnetic fields splitting the transitions. We tried to apply magnetic fields in order to maximize the $T_2$ and thus null any stray fields (assuming $T_2$ is at a maximum at zero field), but we did not observe any significant change in $T_2$ or the echo envelope. Another possibility could be superhyperfine coupling with host yttrium ions. The difference in frequency of the two echo components of B is 74 $\pm$ 4 kHz, which is on the same order as Ref.~\cite{car2018}. We did not experimentally investigate this further.

The main limit to the coherence times in our experiments is the magnetic field fluctuations from $^{167}$Er$^{3+}$-$^{167}$Er$^{3+}$ flip-flops. In Ref.~\cite{rancic2018}, the large field and low temperature helped to obtain the long coherence time by freezing the Er spins. Decreasing the concentration of Er$^{3+}$ ions can also reduce the number of Er-Er flip-flops \cite{bottger2006c}. In a previously reported measurement on transitions A using coherent Raman beats \cite{hashimoto2016a}, reducing the Er$^{3+}$ ion concentration by a factor of 5 resulted in an increase in coherence time by approximately a factor of 4. In principle, the Er$^{3+}$ ion concentration can be reduced to extend the coherence time until the neighboring yttrium ions become the dominant source of dephasing instead, but at the expense of reduced atom-cavity coupling in the case of a microwave memory.

\section{Suitability for a microwave memory}

Because reaching the strong coupling regime is a requirement for an efficient microwave memory, the transition strengths are an important consideration. These can be calculated from the spin Hamiltonian \cite{mcauslan2012}. The resulting strength is 440\,MHz/T for transitions A (calculated using Ref. \cite{chen2018}), which is much larger than the nuclear magneton (7.6\,MHz/T). When compared to electron spin transitions in erbium that have been used to achieve strong coupling \cite{tkalcec2014}, the strengths of transitions A are 35 times smaller than the Bohr magneton but the inhomogeneous broadening is only 350\,kHz, which is 62 times smaller. This means it will be very feasible to reach the strong coupling regime with the sample cooled to the millikelvin regime.

For B, the transition strengths are 60\,MHz/T (calculated using Ref. \cite{horvath2019}). For both A and B, the transition strengths are much larger than that of the nuclear spinlike transition at high field from Ref~\cite{rancic2018}, which is 2.3\,MHz/T. 

The transitions that we have looked at have transition frequencies ($\approx 800\,$MHz) lower than is common to use for superconducting qubits ($\approx 5-10$\,GHz). Frequencies above 5\,GHz are usually used to avoid unwanted thermal excitation. However operating superconducting qubits at the frequencies we use here is possible and sometimes even desirable. The fluxonium qubit of Ref. \cite{nesterov2018}, for example, operates at 500\,MHz.

There are many levels in \pErYSO\ from 0 to 5 GHz, and likewise for site 2 which we did not investigate, so there could also be some flexibility in choice of transition frequency.

\section{Conclusion}

To conclude, we have used Raman heterodyne spectroscopy to measure hyperfine spectra from the ground and excited states of \pErYSO, as well as to detect spin echoes from both states. Our measured coherence times at zero field are approximately 100 times longer than previous measurements for the same Er$^{3+}$ concentration at zero field \cite{hashimoto2016a} and 200 times longer than the coherence time used to demonstrate a microwave memory using the even isotopes of Er$^{3+}$ \cite{probst2015a}. This demonstrates that \pErYSO\ is a potential candidate for a telecom compatible and superconducting qubit compatible microwave quantum memory.

\section{Acknowledgements}

This work was supported by the Marsden Fund of the Royal Society of New Zealand through Contract No. UOO1520

%

\end{document}